\documentclass[pra, showpacs, twocolumn, floatfix]{revtex4}
\usepackage{graphicx}
\usepackage{amsmath, amsfonts, amssymb, bm}
\usepackage[]{psfrag}
\psfragscanoff 

\usepackage{verbatim}
\usepackage{bm}
\begin{document}

\newcommand{\ket}[1]{| #1 \rangle}
\newcommand{\bra}[1]{\langle #1 |}
\title{Collective coherent population trapping in a thermal field}
\author{M. \surname{Macovei}$^{a \star}$}
\author{Z. \surname{Ficek}$^{a,b} $}
\email{ficek@physics.uq.edu.au}
\author{C. H. \surname{Keitel}$^{a} $}
\email{keitel@mpi-hd.mpg.de}
\affiliation{$^{(a)}$Max-Planck Institute for Nuclear Physics, \\
Saupfercheckweg 1, D-69117 Heidelberg, Germany \\
$^{(b)}$Department of Physics, School of Physical Sciences, \\
The University of Queensland, Brisbane, Australia 4072}
\date{\today}
\begin{abstract}
We analyzed the efficiency of coherent population trapping (CPT) in a superposition of the ground states of three-level atoms under the influence of the decoherence process induced by a broadband thermal field. We showed that in a single atom there is no perfect CPT when the atomic transitions are affected by the thermal field. The perfect CPT may occur when only one of the two atomic transitions is affected by the thermal field. In the case when both atomic transitions are affected by the thermal field, we demonstrated that regardless of the intensity of the thermal field the destructive effect on the CPT can be circumvented by the collective behavior of the atoms. An analytic expression was obtained for the populations of the upper atomic levels which can be considered as a measure of the level of thermal decoherence. The results show that the collective interaction between the atoms can significantly enhance the population trapping in that the population of the upper state decreases with increased number of atoms. The physical origin of this feature was explained by the semiclassical dressed atom model of the system. We introduced the concept of multiatom collective coherent population trapping by demonstrating the existence of collective (entangled) states whose storage capacity is larger than that of the equivalent states of independent atoms.
\end{abstract}
\pacs{32.80.Qk, 42.50.Fx, 42.50.Gy }
\maketitle
\section{Introduction}
The study of atomic coherence effects in multilevel atoms is one of the most active area in atomic spectroscopy~\cite{ao76,ari90,fs}. Especially, the theory of coherent population trapping (CPT) in a three-level $\Lambda$-type atom has been extensively studied and the phenomenon has been observed experimentally in a sodium vapor~\cite{alz76,gra78}, photoassociation systems~\cite{dumke}, BEC~\cite{win05} and solids~\cite{kol05}. The CPT results from the formation of a coherent superposition of the ground atomic states that is decoupled from the external fields and hence referred to as a dark state. The particular interest of this phenomenon consists of the possibility of storage and coherent manipulation of the population in a coherent superposition of the ground states of the atoms~\cite{fl00,lu03}. These phenomena have received greatly increased experimental attention in recent years and experimental techniques have been developed which allow a reversible transfer of quantum information from light to the dark state of the atoms~\cite{jsc04}. The coherent population trapping has also been investigated in the context of lasing without inversion~\cite{ha97}, subrecoil laser cooling~\cite{as88} and a search for materials that display a high index of refraction accompanied by vanishing absorption~\cite{mos,sz92,zms}.

The atomic coherence effects are sensitive to decoherence. In the CPT effect, one source of decoherence is fluctuations of the laser fields used to create the coherent superposition of the atomic ground states~\cite{dk82}. The fluctuations redistribute the population among the atomic states including the excited atomic states from which it can be spontaneously emitted resulting in optical losses.
Recent investigations of decoherence processes in atomic systems have demonstrated that CPT and quantum storage in an ensemble of noninteracting atoms are limited primarily by different decoherence processes such as atomic collisions, atom loss and motion of atoms~\cite{dumke,mf05}. The results show an interesting property that in the limit of the total number of excitations much smaller than the number of atoms, the decoherence rate of the multiatom system is of the same order of magnitude as in the single atom, i.e. is independent of the number of atoms in the sample. In an earlier study, Jyotsna and Agarwal~\cite{ja96} showed that the CPT effect in a dense atomic medium is unaffected by local-field effects. 

It is well known that the dominant contribution to the decoherence processes in the interaction of atoms with the electromagnetic field stems from the thermal fluctuations. They are present in a non-zero temperature reservoir to which the atoms are coupled. The fluctuations cause a pumping of the population stored in the dark state into the excited states of the atoms from which it can be spontaneously emitted resulting in an increase in decoherence. The magnitude of thermal fluctuations depends on temperature of the reservoir and determines the minimum level of thermal decoherence.

In this paper we propose a method to suppress the decoherences that occur due to the thermal fluctuations of the environmental electromagnetic reservoir at temperature $T$. Essentially, we examine the CPT effect in three-level $\Lambda$ systems by addressing a practical question: How can one increase the efficiency of trapping and storage of the population in the presence of thermal decoherence. In particular, we will investigate limits to the efficiency of the CPT effect in a single atom and next will explore the role of multiatom collective behavior in the reduction of the single-atom decoherence rate induced by the thermal field. The dipole-dipole interactions between the atoms will not be taken into account here assuming lower atomic densities, so that the collective behavior we consider stems entirely from the mutual coupling of all the atoms with the common radiation field \cite{ex_L}. Employing the analytic solution for the density operator of the system, we find that in general the single-atom coherent population trapping effect, reduced by thermal fluctuations, can be significantly improved or even completely restored when the atoms interact collectively with the thermal modes of the reservoir. We are particularly interested in the manner in which multiatom effects can lead to a suppression of thermal decoherence. With appropriate selection of atomic parameters, we will find cases of almost perfect coherent population trapping in the presence of the thermal decoherence. Our physical interpretation of the results is based on the semiclassical dressed atom model of the collective atomic system. The collective dressed states of the system are identified, and the effect of suppression of the thermal decoherence is explained in terms of the increased capacity of these states. This is shown to arise from correlation-enhanced transition rates among the multiatom dressed states, in particular those entering the trapped state. Hence, the effects of decoherence by thermal fields may by reverted more rapidly.
\section{Approach}
\begin{figure}[b]
\includegraphics[height=3cm]{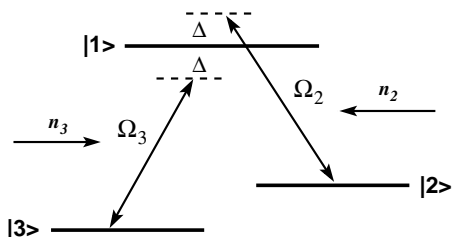}
\caption{\label{fig-1} Energy-level diagram of a three-level $\Lambda$-type atom driven by two laser 
fields of Rabi frequencies $2\Omega_{2}$ and $2\Omega_{3}$.}
\end{figure}

The system we consider is an ensemble of $N$ identical three-level $\Lambda$-type atoms each with excited 
state $|1\rangle$ and two nondegenerate ground states $|2\rangle$ and $|3\rangle$. The atoms are driven by two single-mode 
cw laser fields of Rabi frequencies $2\Omega_{2}$ and $2\Omega_{3}$ and angular frequencies $\omega_{L2}$ and
$\omega_{L3}$ significantly different from each other, so that each laser is coupled only to one of the allowed transitions, as shown in~Fig.~\ref{fig-1}.
The transitions are associated with nonzero dipole moments $\vec{\mu}_{12}$ and $\vec{\mu}_{13}$, and the laser fields 
are detuned from the atomic transition frequencies, such that there is a nonzero two-photon detuning 
$\Delta = (\omega_{13} -\omega_{12}+\omega_{L2} -\omega_{L3})/2$. 
The transition $\ket 2 \rightarrow \ket 3$ is forbidden in the electric dipole approximation 
$(\vec{\mu}_{23}=0)$. The exited atoms may decay spontaneously due to the zero point fluctuations of the electromagnetic 
field from the state $|1\rangle$ to both ground states $|2\rangle$ and $|3\rangle$ with the decay rates $2\gamma_{2}$ and 
$2\gamma_{3}$, respectively. We assume that the atoms are contained in a volume with linear dimensions that are small
compared with the radiation wavelengths, the Dicke model~\cite{dik}. Thus, all atoms experience the same Rabi frequencies 
of the driving fields including their phases, and propagation effects are negligible due to the small size of the sample. 
In addition, we assume that the atomic transitions are driven by a thermal field of the mean photon numbers $\bar n_{2}$ and 
$\bar n_{3}$ at the atomic transition frequencies $\omega_{12}$ and $\omega_{13}$, respectively.

The system is described by the reduced density operator, which in the interaction picture and under the usual Born-Markov 
and rotating-wave approximations satisfies the master equation
\begin{eqnarray}
\frac{\partial \rho}{\partial t} = -\frac{i}{\hbar}[H_{0},\rho]  
+\gamma _{2}{\cal L}_{2}\rho +\gamma _{3}{\cal L}_{3}\rho ,\label{master} 
\end{eqnarray}
where
\begin{eqnarray}
H_{0} &=& \hbar\Delta (S_{22}-S_{33})+\hbar\sum_{\alpha \in \{2,3\}}\Omega_{\alpha} (S_{1\alpha} + S_{\alpha 1}), \nonumber \\
{\cal L}_{2}\rho &=& (1+\bar{n}_{2})[S_{21}\rho ,S_{12}] + \bar{n}_{2} [S_{12}\rho ,S_{21}] + {\rm H.c.} , \nonumber \\
{\cal L}_{3}\rho &=& (1+\bar{n}_{3})[S_{31}\rho ,S_{13}] + \bar{n}_{3} [S_{13}\rho ,S_{31}] + {\rm H.c.} .  
\end{eqnarray}
Here ${\cal L}_{2}\rho$ and ${\cal L}_{3}\rho$ are operators representing the damping of the atoms via spontaneous emission 
and $H_{0}$ is the Hamiltonian describing the coupling of the atoms to the laser fields. The operators $S_{\alpha \beta }$ are the collective atomic operators 
\begin{eqnarray}
S_{\alpha \beta } = \sum_{j=1}^{N}S_{\alpha \beta }^{(j)} = \sum_{j=1}^{N}|\alpha \rangle_{j}{}_{j}\langle \beta | ,
\quad \alpha,\beta = 1,2,3 ,
\end{eqnarray}
which obey the usual commutation relations 
\begin{eqnarray}
[S_{\alpha \beta },S_{\alpha ^{^{\prime }}\beta^{^{\prime }}}] = 
\delta _{\beta \alpha ^{^{\prime }}}S_{\alpha \beta ^{^{\prime }}} - 
\delta _{\beta ^{^{\prime }}\alpha }S_{\alpha ^{^{\prime }}\beta } .\label{com}
\end{eqnarray}
The master equation (\ref{master}) allows to obtain equations of motion for the expectation value of an arbitrary
combination of the atomic operators. The calculations can be performed without much troubles for the simple case of 
a single atom $(N=1)$ and arbitrary $\Delta$. However, for $N>1$ the calculation of the expectation value is not an 
easy task. In even the simplest cases of small numbers of atoms, the calculations are prohibitively difficult due to the enormity of the number of coupled equation of motion. Fortunately, for the $\Delta =0$ case and high field strengths, $\Omega_{k}\gg N\gamma_{k}$, an approximation technique has been developed, which greatly simplifies the master equation (\ref{master}) and thus ables to perform analytical calculations of the expectation value of an arbitrary combination of the atomic operators. The restriction to the $\Delta =0$ case stems from the difficulty in obtaining a closed set of equations when the two-photon detuning is present~\cite{cor82}. A full discussion of the technique is 
given in Refs.~\cite{symm,ag73,at76,ag78,pu94,law,bog,mek}. In the interest of brevity only the key results will be given here.
The technique is implemented by introducing dressed states of a single atom, which are obtained by a diagonalization 
of the single-atom interaction Hamiltonian 
\begin{eqnarray}
H_{0j} = \Omega_{2} (S_{12}^{(j)}+S_{21}^{(j)}) + \Omega_{3} (S_{13}^{(j)}+S_{31}^{(j)}) .  
\end{eqnarray}
The single-atom dressed states are of the form
\begin{eqnarray}
\ket{\Psi_{1}}_{j} &=& \frac{1}{\Omega}\left(\Omega_{2}\ket 3_{j} -\Omega_{3}\ket 2_{j}\right) , \nonumber \\
\ket{\Psi_{2}}_{j} &=& \frac{1}{\sqrt{2}}\ket 1_{j} +\frac{1}{\sqrt{2}\Omega}\left(\Omega_{2}\ket{2}_{j} +\Omega_{3}\ket{3}_{j}\right) , \nonumber \\
\ket{\Psi_{3}}_{j} &=& \frac{1}{\sqrt{2}}\ket 1_{j} -\frac{1}{\sqrt{2}\Omega}\left(\Omega_{2}\ket{2}_{j} +\Omega_{3}\ket{3}_{j}\right) , \label{DS}
\end{eqnarray}
where  $\Omega =\sqrt{\Omega^{2}_{2}+\Omega^{3}_{3}}$ is the generalized Rabi frequency.

The idea of the approximate technique is now to replace the collective operators $S_{\alpha \beta }$ by the 
collective dressed-atom operators
\begin{eqnarray}
R_{\alpha \beta} = \sum^{N}_{j=1}R^{(j)}_{\alpha \beta} = \sum^{N}_{j=1}|\Psi_{\alpha} 
\rangle_{j}{}_{j}\langle \Psi_{\beta}| , \quad \alpha,\beta = 1,2,3 ,\label{rab}
\end{eqnarray}
and then substitute for $S_{\alpha \beta }$ into the damping terms of the master equation (\ref{master}). Next, we make 
the unitary transformation of the density operator
\begin{eqnarray}
\tilde{\rho} = \exp\left(\frac{i}{\hbar}\tilde{H}_{0}t\right)\rho \exp\left(-\frac{i}{\hbar}\tilde{H}_{0}t\right) ,
\end{eqnarray}
where
\begin{eqnarray}
\tilde{H}_{0} = \hbar\Omega \left(R_{22} - R_{33}\right) = \hbar\Omega R_{z}
\end{eqnarray}
and on carrying out this procedure it is found that certain terms in the transformed master equation are slowly varying
while the others are rapidly oscillating at frequencies $\Omega$ and $2\Omega$. The approximation then consists of dropping 
these rapidly oscillating terms. The master equation (\ref{master}) in the dressed state basis reduces to
\begin{eqnarray}
\frac{\partial \tilde{\rho}}{\partial t} &=& -i\Omega[R_{z},\tilde{\rho}] + \{ \Gamma _{0}([R_{z}\tilde{\rho}, R_{z}] + [R_{32}\tilde{\rho},R_{23}] \nonumber \\
&+&[R_{23}\tilde{\rho}, R_{32}]) + \Gamma_{1}([R_{12}\tilde{\rho}, R_{21}]+[R_{13}\tilde{\rho}, R_{31}]) \nonumber \\
&+& \Gamma_{2}([R_{21}\tilde{\rho},R_{12}] + [R_{31}\tilde{\rho}, R_{13}]) + {\rm H.c.} \}, \label{masterd} 
\end{eqnarray}
where
\begin{eqnarray} 
\Gamma_{0} &=&\frac{1}{2}\{\gamma_{2}(1 + 2\bar n_{2})[\frac{\Omega_{2}}{\sqrt{2}\Omega}]^{2} + \gamma_{3}(1 + 2\bar n_{3})[\frac{\Omega_{3}}{\sqrt{2}\Omega}]^{2}\},\nonumber \\
\Gamma_{1} &=& \frac{1}{2}\{\gamma_{2}(1 + \bar n_{2})[\Omega_{3}/\Omega]^{2} + \gamma_{3}(1 + \bar n_{3})[\Omega_{2}/\Omega]^{2}\},\nonumber \\
\Gamma_{2} &=& \frac{1}{2}\{\gamma_{2}\bar n_{2}[\Omega_{3}/\Omega]^{2} + \gamma_{3}\bar n_{3}[\Omega_{2}/\Omega]^{2}\},\label{trd}
\end{eqnarray}
are the transition rates between the single-atom dressed states.

Using the approximate master equation, it is straightforward to obtain a simple analytical solution for the steady-state 
density operator of the system. The solution can be written in the form
\begin{eqnarray}
\rho_{s} = Z^{-1}\exp[-\xi R_{11}] , \label{SS} 
\end{eqnarray}
where
\begin{eqnarray} 
\xi = \ln\left[\frac{\Omega^{2}_{3}\bar n_{2} 
+ \eta \Omega^{2}_{2}\bar n_{3}}{\Omega^{2}_{3}(1+\bar n_{2}) 
+ \eta \Omega^{2}_{2}(1+\bar n_{3})}\right] ,\label{xi}
\end{eqnarray}
and $\eta =\gamma_{3}/\gamma_{2}$. The parameter $Z$ is the normalization constant such that Tr$\{\rho_{s}\}=1$. It is easily verified that~$\xi$ is always negative independent of the parameters used and approaches zero when $\bar n_{2}$ and/or $\bar n_{3}$ go to infinity. The solution~(\ref{SS}) was obtained in Refs.~\cite{bog,law,mek}, and some applications are discussed there in details. In Ref.~\cite{mek}, the solution has been used to investigate different control schemes for collective systems of three-level atoms. In this paper, we focus on the competition between thermal fluctuations and the collective effects that can lead to collective population trapping.

The steady-state solution (\ref{SS}) enables to calculate any statistical moment of the diagonal elements 
$R_{\alpha \alpha}$, and thus population distributions between atomic states.
In particular, an $k$-th order moment of $R_{11}$ (expectation value of a product of $k$ 
operators $R_{11}$), is of the form 
\begin{eqnarray}
\langle R^{k}_{11}\rangle_{s} = (-1)^{k}Z^{-1}\frac{\partial^{k}}{\partial \xi^{k}}Z , \quad k=1,2,\ldots ,\label{r1sol}
\end{eqnarray}
and the first order statistical moments of $R_{22}$ and $R_{33}$~are
\begin{eqnarray}
\langle R_{22}\rangle_{s} = \langle R_{33}\rangle_{s} = [N - \langle R_{11}\rangle_{s}]/2 ,\label{rsol}
\end{eqnarray}
where 
\begin{eqnarray}
Z=\frac{N+2-(N+1){\rm e}^{\xi}-{\rm e}^{-\xi(N+1)}}{(1-{\rm e}^{\xi})(1-{\rm e}^{-\xi})} . \label{Z}
\end{eqnarray}  
One can easily show from (\ref{SS}) that the steady-state off-diagonal elements $R_{\alpha \beta}\ (\alpha\neq \beta)$ 
equal zero. Note from Eq.~(\ref{rsol}) that all the non-zero expectation values can be represented in terms of 
$\langle R^{k}_{11}\rangle_{s}$. The steady-state solutions are to be used in the forthcoming treatment of the coherent 
population trapping in a multiatom system. 
\section{Coherent Population Trapping}
Before we proceed to the detailed analysis of the multiatom trapping effect, we briefly investigate the trapping behavior of single atoms in the presence of thermal fluctuations. In this way we may see what restrictions are brought
by the thermal fluctuations for the trapping phenomenon and how they are related to the coherent driving process.
Coherent population trapping effect in a system of three-level atoms may be monitored experimentally in terms of the 
intensity of the fluorescence light emitted~\cite{alz76,gra78,dumke,win05,kol05}. It is manifested by the disappearance of the fluorescence which, on the other side, manifests the vanishing of the population of the upper atomic states~$\ket 1_{j}$. Therefore, we will consider first the effect of the thermal field on the so-called transparency window, i.e. the dependence of the stationary population $\rho_{11}^{s}$ on the two-photon detuning. Next, using the stationary solution (\ref{SS}), we will find the analytical expression for the population at the two-photon resonance, $\Delta =0$, and will analyze how one could reduce the destructive effect of the thermal field on the minimum of the population at $\Delta =0$.
\begin{figure}[t]
\includegraphics[height=4cm]{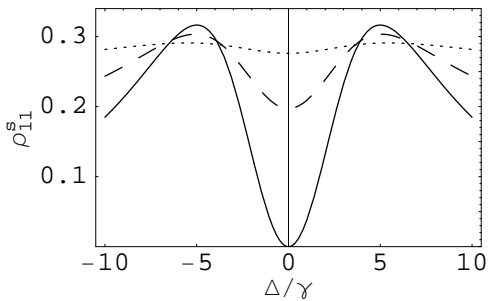}
\caption{\label{fig-2} Stationary population of the upper state $\ket 1_{j}$ as a function of the two-photon detuning 
$\Delta$ for $\gamma_{2}=\gamma_{3}=\gamma $, $\Omega_{2}=\Omega_{3}=5\gamma$ and different $\bar{n}$: $\bar{n}=0$ (solid line), $\bar{n}=0.5$ (dashed line), $\bar{n}=2$ (short dashed line).}
\end{figure} 

Fig.~(\ref{fig-2}) illustrates the stationary population $\rho_{11}^{s}$ as a function of the two-photon detuning $\Delta$. We have obtained the population by solving numerically the master equation (\ref{master}) for $N=1$. 
It is seen that in the absence of the thermal field, $\bar{n}=0$, there is perfect CPT observed at $\Delta =0$. When the atom is in the thermal field equally affecting both transitions, the CPT effect is reduced and the thermal field washes out the transparency window as $\bar{n}\gg 1$. Thus, the thermal field has a destructive effect on the CPT, because the thermal field is an incoherent field with random fluctuations that destroy the coherent process induced by the laser fields. 

The variation with $\bar{n}$ of the minimum of the upper state population at $\Delta =0$ can be analyzed explicitly 
using Eq.~(\ref{DS}) which for~$N=1$ and together with the steady-state solution (\ref{SS}) gives a simple analytical 
expression for $\rho^{s}_{11}$ in the form 
\begin{eqnarray}
\rho_{11}^{s} = \frac{1}{2}\left(\langle R_{22}\rangle_{s} + \langle R_{33}\rangle_{s}\right)
= \frac{{\rm e}^{\xi}}{1+2{\rm e}^{\xi}} . \label{UP}
\end{eqnarray}

First, we note from Eq.~(\ref{UP}) that the population distribution between the atomic states
is determined solely by the parameter~$\xi$. Clearly, the population distribution and consequently the trapping effect depend on several 
parameters such as the laser intensity, spontaneous emission rates, and mean number of thermal photons.
We can call the parameter $\xi$ as a measure of efficiency of the CPT effect. 

Here the efficiency of the CPT is examined in various intensity regimes of the coherent fields for equal and also unequal 
average numbers of thermal photons. The average numbers can be made unequal by a suitable choice of bandwidth of the thermal field. 
The selective excitation of the atomic transitions can be realized in practice by applying a finite bandwidth multimode thermal field whose 
bandwidth is much smaller than the splitting of the lower atomic levels, but large compared to the natural linewidths of the 
atomic transitions to satisfy the Markov approximation used in the derivation of the master equation.

In the limit of $\bar n_{2}=0$ that the thermal fluctuations affect only the $\ket 1 \rightarrow \ket 3$ transition, the
parameter $\xi$ reduces to
\begin{eqnarray} 
\xi = \ln\left[\frac{\eta\Omega^{2}_{2}\bar n_{3}}{\Omega^{2}_{3} 
+ \eta \Omega^{2}_{2}(1+\bar n_{3})}\right] .\label{xi1}
\end{eqnarray}
The parameter $\xi$ does not change substantially with the Rabi frequencies unless $\Omega_{3}$ is much larger than the
Rabi frequency $\Omega_{2}$ of the other transition. In the very strong-field
regime of $\Omega^{2}_{3}\gg \eta \Omega^{2}_{2}(1+\bar n_{3})$, the parameter $\xi$ approaches the limit of
$\xi\rightarrow -\infty$. This minimum value is that one which leads to vanishing of the population of the 
upper atomic state, because $\lim_{\xi \to -\infty}\rho^{s}_{11} =0 $. This predicts that perfect coherent population 
trapping can be observed even in the presence of thermal decoherence on one of the two atomic transitions, which is in contrast to the result of~\cite{bla97}. However, it requires that the transition influenced by the decoherence is simultaneously driven by a strong laser field. It can be understood rather easily. For a large Rabi frequency $\Omega_{3}$, the coherent processes on the $\ket 1 \rightarrow \ket 3$ transition dominate over the incoherent thermal processes resulting in perfect transparency.

Various other intensity regimes can also be distinguished. If $\bar n_{2}\neq \bar n_{3}$, the parameter $\xi$ 
can depend entirely on $\bar n_{2}$ or $\bar n_{3}$ depending on the ratio $\Omega_{3}/\Omega_{2}$. For instance, 
when $\bar{n}_{2}\Omega^{2}_{3}\gg \eta \bar{n}_{3}\Omega^{2}_{2}$, we find that
\begin{eqnarray} 
\xi = \ln\left(\frac{\bar n_{2}}{1+\bar n_{2}}\right) .\label{xi2}
\end{eqnarray}
This predicts that the coherent population trapping depends entirely on the thermal fluctuations at the weakly driven 
$\ket 1 \rightarrow \ket 2$ transition. In the opposite limit of $\eta\bar{n}_{3}\Omega^{2}_{2}\gg \bar{n}_{2}\Omega^{2}_{3}$, the parameter $\xi$ now depends entirely on $\bar n_{3}$. Thus, the driving fields are relatively efficient in controlling decoherence in a single atom. Again, it can be interpreted as caused by coherent processes that dominate
incoherent thermal processes on the strongly driven transition. This also shows that the suppression of the thermal
decoherence in a single atom is limited to the level set by the lowest thermal fluctuations affecting the atomic transitions.

\begin{figure}[t]
\includegraphics[height=4cm]{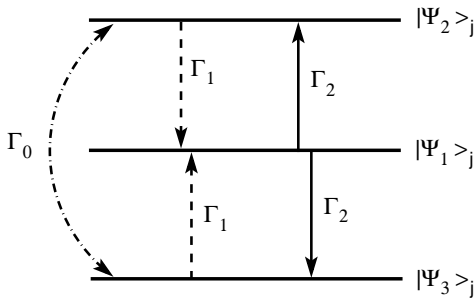}
\caption{\label{fig-3} Single atom dressed states and possible transitions with the rates $\Gamma_{0}, \Gamma_{1}$
and $\Gamma_{2}$.}
\end{figure}

In the case when the thermal field equally contributes to both atomic transitions, $\bar n_{2}=\bar n_{3}\equiv\bar n$, 
we have
\begin{eqnarray} 
\xi = \ln\left(\frac{\bar n}{1+\bar n}\right) ,\label{xi3}
\end{eqnarray}
independent of the Rabi frequencies and the spontaneous emission rates. Obviously, the trapping effect is reduced 
regardless of how strong are the Rabi frequencies of the laser fields relative to the thermal fluctuations.
In other words, there is no possibility of obtaining perfect population trapping or a control of the decoherence level 
in a single atom when both transitions are equally affected by the thermal field. 
A qualitative understanding of this effect can be obtained in terms of the transition rates (\ref{trd}).  Figure~\ref{fig-3} shows the single-atom dressed states and the transition rates $\Gamma$. One can see from the figure that the population flows into the state $\ket{\Psi_{1}}_{j}$ with the rate $\Gamma_{1}$, and is removed from this state with the rate $\Gamma_{2}$. The state $\ket{\Psi_{1}}_{j}$ is a linear superposition of only the ground states of the atom that it is the trapping (dark) state. Therefore, we can call the rate $\Gamma_{2}$ a decoherence rate, as it transfers the population from the dark state to the upper state $\ket 1$ from which it can be spontaneously radiated resulting in an increase in decoherence and optical losses. Only in the absence of the thermal field, $\bar n_{2}=\bar n_{3}=0$, the transition rate $\Gamma_{2}=0$. 
Evidently, the CPT effect depends crucially on $\Gamma_{2}$, and therefore the key to maintain a large efficiency of the CPT is to make $\Gamma_{2}$ as small as possible. 
It can be done when the thermal field unequally affects the atomic transitions, i.e. when the number of thermal photons affecting one of the transitions is different than on the other transition. 
For example, when $\bar n_{2}\ll \bar n_{3}$, the rate $\Gamma_{2}$ can be made small, proportional to $\bar n_{2}$, by changing the ratio $\Omega_{3}/\Omega_{2}$. It is easily to see from (\ref{trd}) that in the case of $\Omega_{3}\gg \Omega_{2}$, the rate $\Gamma_{2}$ is only of the order of $\bar n_{2}$ despite the fact that there is a large number of thermal photons present on the $\ket 1 -\ket 3$ transition. When the thermal field equally contributes to both atomic transitions, $\gamma_{2}\bar n_{2}=\gamma_{3}\bar n_{3}=\gamma\bar{n}$, and from Eq.~(\ref{trd}) we find that $\Gamma_{2}=\bar{n}\gamma/2$ independent of the Rabi frequencies of the laser fields. This is the smallest decoherence rate one can achieve in the single atom interacting with a thermal field that equally affects the atomic transitions. The limit is set by the number of photons $\bar{n}$ that, on the other hand, depends on temperature of the reservoir. An improvement of the CPT effect in the $\Lambda-$ type system with asymmetric spontaneous decay rates has been predicted in the absence of the thermal field \cite{PRA52}, but in this case the transparency window shows a strong sensitivity to the Rabi frequencies and is observed only in the limit of very weak driving fields.

The limit set in single atoms by temperature of the reservoir can be circumvented to improve the efficiency of the CPT effect if one considers multiatom collective systems in which interatomic interactions can create collective states of a significantly enhanced storage capacity compared with the capacity of the corresponding states of individual atoms.
\section{Collective trapping states}
The effects described in Section III can be seen in dilute atomic gases where the interatomic interactions are not important.
However, a more interesting situation emerges as we have considered here atomic samples where radiative interactions between 
the atoms can lead to a collective (entangled) behavior of the atoms. Here, we include the multiatom effects and calculate the 
population $\rho_{11}^{s}$ as a function of the number of atoms and the number of thermal photons.

The upper state population $\rho^{s}_{11}$ can be evaluated using Eq.~(\ref{DS}) which, together with the steady-state solution 
(\ref{SS}), gives the analytical expression for $\rho^{s}_{11}$ in terms of $\xi$ and $N$ as  
\begin{eqnarray}
\rho_{11}^{s} &=& \frac{Z^{-1}}{\left(1-{\rm e}^{-\xi}\right)}[\frac{1}{2}N(N + 1) \nonumber \\
&-& \frac{{\rm e}^{-\xi N} + N{\rm e}^{\xi} - N - 1}{(1 - {\rm e}^{\xi})^{2}}]. \label{CUP}
\end{eqnarray}

\begin{figure}[b]
\includegraphics[width = 7cm]{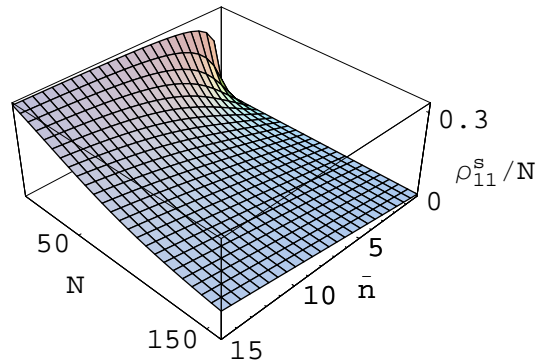}
\caption{\label{fig-4} The upper-state population $\rho_{11}^{s}/N$ as a function of $\bar n$ and $N$ 
for $\bar n_{2} =\bar n_{3} =\bar n$ and moderate numbers of atoms.}
\end{figure}

In Fig.~\ref{fig-4}, we present a three-dimensional plot which shows that in the absence of the thermal field, i.e. 
$\bar n =0$, the stationary population $\rho_{11}^{s}$ is equal to zero independent of the number of atoms.
Thus, for $\bar n =0$ the collective behavior of the atoms does not affect the trapping effect. The presence of the 
thermal field has a destructive effect on the trapping phenomenon that the population in the upper state is no longer
zero and increases with increasing number of thermal photons $\bar n$. However, the rate of the increase of the population
decreases with increasing number of atoms $N$ such that for a suitably large $N$ the population $\rho_{11}^{s}$ 
may remain very small even for large $\bar n$. In other words, the thermal decoherence decreases with increasing number of atoms. 
Thus, the collective interactions are relatively efficient in suppression of thermal decoherence such that the atoms may remain 
in their ground states even in the presence of the thermal decoherence. This is a suprising result as one might expect that 
decoherence should increase with the increasing number of atoms. 

\begin{figure}[t]
\includegraphics[height=4cm]{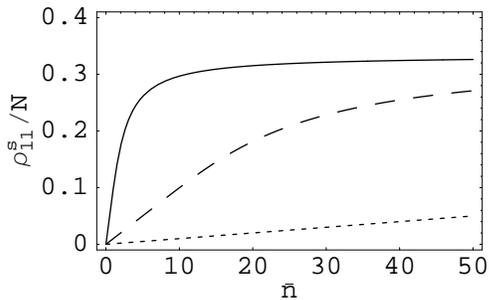}
\caption{\label{fig-5} The upper-state population $\rho_{11}^{s}/N$ as a function of $\bar n_{2}=\bar n_{3}=\bar n$ for different 
numbers of atoms: $N=10$ (solid line), $N=100$ (dashed line), $N=1000$ (short dashed line).}
\end{figure}

Figure~\ref{fig-5} shows the population $\rho_{11}^{s}$ as a function of $\bar n_{2}=\bar n_{3} \equiv \bar n$ for different numbers of atoms.
Here we see that the rate of the increase of the population decreases with $N$. For a small number of atoms, the population saturates quickly 
with $\bar n$. But, for a large number of atoms, a much stronger thermal field is required to reach saturation. In other words, the collective 
population stored in the ground states is less affected by the thermal fluctuations than for the case of independent atoms. As a consequence, one has a 
practical scheme to reduce thermal decoherence and preserve CPT in the thermal field.

In order to obtain an insight into the physical origin of the reduction of thermal decoherence and the improvement of the population trapping,
we examine the energy structure of the collective system. In general, in the absence of the driving fields, the system can be represented in 
terms of collective symmetric and antisymmetric states. However, in the case of $N$ identical atoms contained in a volume with linear dimensions 
that are small compared with the radiation wavelengths, only the symmetric states couples to external driving fields. The dipole-dipole interactions 
between the atoms lead to a shift of these states from the laser resonance \cite{fs}. Thus, here the Rabi frequencies should be larger than the shift caused by
the dipole-dipole effects, i.e. the latter can be neglected. The antisymmetric states do not participate in the dynamics of the small sample system~\cite{dik}. 
Therefore, we may limit the dynamics to only those involving the symmetric states. Moreover, only the lowest in energy symmetric states are of interest in the 
analysis of the collective population trapping. We therefore consider the lowest energy states defined as 
\begin{widetext}
\begin{eqnarray}
    \ket 3 &=& \left(
\begin{array}{c}
N\\
0
\end{array}
\right)^{-\frac{1}{2}}
\ket{3_{1},3_{2},\ldots,3_{N}} ,\nonumber \\
    \ket 2 &=& \left(
\begin{array}{c}
N\\
1
\end{array}
\right)^{-\frac{1}{2}}
    \sum_{i=1}^{N} \ket{3_{1},\ldots,2_{i},\ldots,3_{N}} ,\nonumber \\
    \ket 1 &=& \left(
\begin{array}{c}
N\\
1
\end{array}
\right)^{-\frac{1}{2}}
    \sum_{i=1}^{N} 
    \ket{3_{1},\ldots,1_{i},\ldots,3_{N}} ,\nonumber \\
    \ket{2^{2}} &\equiv& \ket{22} = \left(
\begin{array}{c}
N\\
2
\end{array}
\right)^{-\frac{1}{2}} \sum_{i<j=1}^{N} 
    \ket{3_{1},\ldots,2_{i},\ldots,2_{j},\ldots,3_{N}} ,\nonumber \\
    \ket{12} &=& \frac{1}{\sqrt{2}}\left(
\begin{array}{c}
N\\
2
\end{array}
\right)^{-\frac{1}{2}}
    \sum_{i\neq j=1}^{N} 
    \ket{3_{1},\ldots,1_{i},\ldots,2_{j},\ldots,3_{N}} ,\nonumber \\
    \ket{2^{3}} &\equiv& \ket{222} = \left(
\begin{array}{c}
N\\
3
\end{array}
\right)^{-\frac{1}{2}} \sum_{i<j<k =1}^{N} 
    \ket{3_{1},\ldots,2_{i},\ldots,2_{j},\ldots,2_{k}\ldots,3_{N}} ,\nonumber \\
    \mathrm{etc.} ,&& \label{4.1}
\end{eqnarray}
\end{widetext}
where the binomial coefficients are the normalization constants. The states (\ref{4.1}) are superpositions of single-atom product 
states $\ket{m}_{i}\otimes \ket{n}_{j}\otimes\ldots\otimes\ket{k}_{l}$ that are symmetric under the exchange of any pair of atoms. 
For example, the state $\ket 2$ is a linear superposition of the product states in which atom $i$ is in the state $\ket 2_{i}$ and 
the remaining $N-1$ atoms are in their states $\ket 3_{j}$. 

\begin{figure}[t]
\includegraphics[width=8cm]{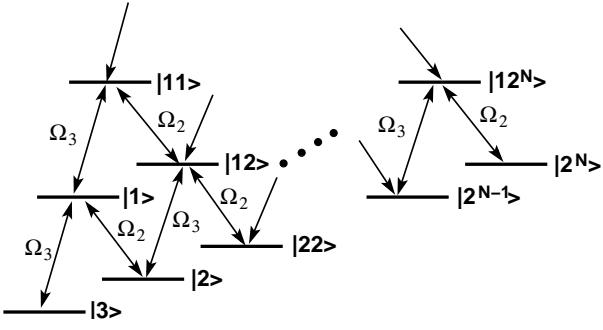}
\caption{\label{fig-6} Energy-level structure of noninteracting collective states of the $N$ atom system.}
\end{figure}

If we now allow the atoms to interact with the laser fields, each state $\ket{2^{k}}$ couples to the first excited states 
$\ket{12^{k-1}}$ and $\ket{2^{k}}$ with the Rabi frequencies $\Omega_{2}$ and $\Omega_{3}$, respectively. Figure~\ref{fig-6} shows 
the collective symmetric states and possible couplings of the two laser fields. As we have already mentioned, we limit the presentation 
to the lowest energy levels which will be mixed together by the interaction leading to a ground collective dressed state, which is of 
the main interest here. The lowest energy state is the product state $\ket 3$=$\ket{3_{1},3_{2},\ldots,3_{N}}$. Each succeeding state 
$\ket{2^{k}}$ is of energy higher by successive increments of $\delta = \omega_{13}-\omega_{12}$. Similarly, each succeeding state 
$\ket{12^{k}}$ is of energy higher by successive increments of $\delta$. It is interesting to note that the rotating-wave approximation, 
which we are assuming to be valid, ignores coupling of states which differ in excitation by two and higher. In other words, the laser 
fields couple only the neighboring ground states through the first excited states. It forms a two-dimensional chain of $\Lambda$ 
configurations.
With the state ordering $\ket 3, \ket 1, \ket 2, \ket{12}, \ket{22}, \ldots ,\ket{2^{N}}$, corresponding to 
the path of successive excitations of the states~$\ket{2^{k}}$ by the laser fields, the interaction Hamiltonian 
$H_{0}$ can be expressed as an infinite tridiagonal matrix 
\begin{widetext}
\begin{eqnarray}
H_{0}/\hbar = \left(
\begin{array}{cccccc}
-N\Delta & \Omega_{3}\sqrt{N} & 0 & 0 & 0 & \cdots\\
\Omega_{3}\sqrt{N} & -(N-1)\Delta & \Omega_{2}\sqrt{1} & 0 & 0 & \cdots\\
0 & \Omega_{2}\sqrt{1} & -N\Delta & \Omega_{3}\sqrt{N-1} & 0 & \cdots\\
0 & 0 & \Omega_{3}\sqrt{N-1} & -(N-1)\Delta & \Omega_{2}\sqrt{2} & \cdots\\
0 & 0 & 0 & \Omega_{2}\sqrt{2} & -N\Delta & \cdots \\
\vdots & \vdots & \vdots & \vdots & \vdots & \ddots
\end{array}
\right) .\label{4.2a}
\end{eqnarray}
\end{widetext}
It is interesting to note that the matrix element of the $\ket 3 -\ket 1$ transition coupled by the Rabi 
frequency $\Omega_{3}$ is enhanced by a factor $\sqrt{N}$ and the magnitude of the matrix elements of the 
successive transitions coupled by the same field decreases along the path to the state $\ket{2^{N}}$.
On the other hand, the matrix element of the $\ket 2 -\ket 1$ transition coupled by the Rabi frequency $\Omega_{2}$
is the same as in the single atom case, but the the magnitude of the matrix elements of the successive transitions 
coupled by the same field increases as $\sqrt{k}$ when one moves along the excitation path to the state $\ket{2^{N}}$.
Thus, the coupling strength of the lasers to the atoms is transferred from one field to the other as one moves 
along the path of excitations from $\ket 3$ to $\ket{2^{N}}$.

We now proceed to diagonalize the matrix (\ref{4.2a}) which will result in collective dressed states. 
The diagonalization is performed by solving Schr\"{o}dinger's time-independent equation in the form
\begin{eqnarray}
\left(H_{0} -\lambda_{n}I\right)\ket{D_{n}}_{N} = 0 ,\label{4.2b}
\end{eqnarray}
where $I$ is the identity matrix and $\ket{D_{n}}_{N}$ is an eigenvector.

Substituting Eq.~(\ref{4.2a}) into Eq.~(\ref{4.2b}) yields the eigenvalue equation
\begin{eqnarray}
&&(\lambda_{n} + N\Delta)\{[\lambda_{n} + (N - 1)\Delta ]( \cdots ) + \Omega_{2}^{2}[\lambda_{n} \nonumber \\
&& + (N - 1)\Delta ] ( \cdots ) \} + \Omega_{3}^{2}N [(\lambda_{n} + N\Delta) (\cdots ) \nonumber \\
&& + \Omega_{3}^{2}(N - 1 )( \cdots )] = 0 \,
\end{eqnarray}
where the $\left(\cdots\right)$ refers to terms of which the explicit form is not needed apart from that those are polynomial functions
of $\lambda_{n}$. It is easily to show that in the case of $\Delta =0$, the eigenvalue equation reduces to
\begin{eqnarray}
    \lambda_{n}\left[\lambda_{n}\left(\cdots\right) + \Omega_{2}^{2}\lambda_{n}\left(\cdots\right) 
    +\Omega_{3}^{2}N\left(\cdots\right)\right] = 0 \,
\end{eqnarray}
from which we see that $\lambda_{n} =0$ is one eigenvalue of $H_{0}$.

In the single atom case the dressed state $\ket{\Psi_{1}}_{j}$ corresponding to the zero eigenvalue was of very
special significance as corresponding to a trapping (dark) state completely decoupled from the fields~\cite{zms,mf05,usb}.
Let us investigate this possibility in the multiatom case.

If we represent the eigenvector $\ket{D_{n}}_{N}$ by the column vector
\begin{eqnarray}
  \ket{D_{n}}_{N}   = 
\left(
\begin{array}{c}
c_{1}\\
c_{2}\\
\vdots\\
c_{i}\\
\vdots\\
c_{n}
\end{array}
\right) ,\label{4.2c}
\end{eqnarray}
we find by substituting into Eq.~(\ref{4.2b}) that for $\lambda_{n}=0$ the coefficients $c_{n}$ for even $n$
are all zero, whereas for odd $n$ the coefficients are given by the recurrence relation
\begin{eqnarray}
c_{2k+1}=\left(-\frac{\Omega_{3}}{\Omega_{2}}\right)^{k}\sqrt{\frac{N!}{k!(N-k)!}}\ c_{1}, ~~k = 1,2,3, \ldots 
\end{eqnarray}
and $c_{1}$ is found from the normalization condition.

The dressed state corresponding to the eigenvalue $\lambda_{n}=0$ can thus be written as 
\begin{eqnarray}
\ket{D}_{N} & \equiv & \ket{D_{0}}_{N} \nonumber \\
&=& \left(\cos\theta\right)^{N} \sum_{k=0}^{N} \left(
\begin{array}{c}
N\\
k
\end{array}
\right)^{\frac{1}{2}} \left(-\tan\theta\right)^{k} \ket{2^{k}} ,\label{4.3}
\end{eqnarray}
where $\ket{2^{0}} = \ket 3$, and
\begin{eqnarray}
    \tan\theta =\frac{\Omega_{3}}{\Omega_{2} } .\label{4.4}
\end{eqnarray}
The collective dressed state (\ref{4.3}) is a linear combination of the state $\ket 3$ and $N$ of the states $\ket{2^{k}}$.
The important feature of the state is that it does not contain the excited states of the atoms
and thus does not radiate. The dressed state is a stationary state of the Hamiltonian $H_{0}$ describing 
the atoms driven by two coherent fields. Therefore, if nothing else is allowed to interact with this system, the state (\ref{4.3}) 
will never change in time. 

We can write the multi-atom dressed state (\ref{4.3}) in the basis of the single-atom dressed states (\ref{DS}). Surprisely, we 
find that the state is of the form
\begin{eqnarray}
\ket{D}_{N} = \ket{\Psi_{1}}_{1}\otimes\ket{\Psi_{1}}_{2}\otimes\cdots\otimes\ket{\Psi_{1}}_{N}, \label{prod}
\end{eqnarray}
which is a product of the single-atom trapping states $\ket{\Psi_{1}}_{j}$. Obviously, the state~(\ref{prod}) is not entangled, which 
shows that trapping of the population in all of the atomic ground states is equally effective in destroying collective (entangled) 
properties of the system. Thus, the improvement of the CPT in the collective multiatom system, seen in Figs.~\ref{fig-4} and~\ref{fig-5}, 
does not arise from collective excitations of the dark state~$\ket{D}_{N}$.

We note in passing that the state in Eq.~(\ref{4.3}) is similar in form to that found by Mewes and Fleischhauer~\cite{mf05},
see also~\cite{fl00,lu03}, who considered collective quantum memories in three-level atoms driven by a classical field and a 
single-mode quantum field. The results of their work demonstrate that the dark states of the multiatom system are highly entangled states. 
However, the state $\ket{D}_{N}$ which is the analog of the dark states found in~\cite{mf05}, is not entangled. Thus, a question arises: 
Why does the dark state $\ket{D}_{N}$ is not entangled? The reason is that the state $\ket{D}_{N}$ is a linear superposition of all the 
collective ground states, whereas the dark states considered in Ref.~\cite{mf05} is restricted to having involved a small number of the 
ground states corresponding to a small number of excitations $k\ll N$. It is easily verified that if we limit the number of the states 
involved in the superposition (\ref{4.3}) to $k<N$, then the resulting state cannot be written as a product of the single atomic states. 
Clearly, entanglement properties of the dark state $\ket{D}_{N}$ depend on the number of atoms involved in the interaction with the laser 
fields, that only for $k<N$ the interaction can produce a dark state which is an entangled state.

To find the explanation why the CPT in the collective system interacting with the thermal field decoheres slower than the system of independent 
atoms, we introduce the interaction between the collective dressed states and the thermal field. This interaction leads to a distribution of 
the population, initially trapped in the dark state $\ket{D}_{N}$, among the collective dressed states. Let us look at the evolution of the 
population of the state $\ket{D}_{N}$. Using the master equation~(\ref{masterd}), we obtain the following equation of motion for the population 
of the state $\ket{D}_{N}$:
\begin{eqnarray}
\dot{\rho}_{DD} = -4N\Gamma_{2}\rho_{DD} + 2N\Gamma_{1}\left(\rho_{D2} + \rho_{D3}\right), \label{eqm}
\end{eqnarray}
where $\rho_{DD}$ is the population of the state $\ket{D}_{N}$ and $\rho_{D2}$ are $\rho_{D3}$ are populations of the following superposition states 
\begin{eqnarray}
\ket{D2}_{N} \equiv \frac{1}{\sqrt{N}}\sum^{N}_{j=1} \ket{\Psi_{1}^{(1)},\Psi_{1}^{(2)},\ldots ,\Psi_{2}^{(j)},\ldots ,\Psi_{1}^{(N)}}, \nonumber \\
\ket{D3}_{N} \equiv \frac{1}{\sqrt{N}}\sum^{N}_{j=1} \ket{\Psi_{1}^{(1)},\Psi_{1}^{(2)},\ldots ,\Psi_{3}^{(j)},\ldots ,\Psi_{1}^{(N)}}, \label{6.1}
\end{eqnarray}
which differ in energy from the state $\ket{D}_{N}$ by $+\Omega$ and $-\Omega$, respectively. 

The states $\ket{D2}_{N}$ and $\ket{D3}_{N}$ are linear superpositions of the product states in which $N-1$ atoms are in state $\ket{\Psi_{1}}_{j}$ 
and one atom is in the state $\ket{\Psi_{2}}_{j}$ and $\ket{\Psi_{3}}_{j}$, respectively.
It is interesting to note from Eq.~(\ref{eqm}) that both spontaneous emission and the thermal field couple the state $\ket{D}_{N}$ to only those states 
which differ in the excitation by one. Moreover, the transition rates between these states are $N$ times larger than that for single atoms. This shows 
that the system is superradiant despite the fact that the state $\ket{D}_{N}$ is the product state of the single-atom trapping states. In addition, the 
collective decay rate of the radiators entering the state $\ket{D}_{N}$ is larger than that describing the atoms escaping from it. Thus, the collective 
properties of the system are preserved due to the presence of the superposition states involving the single-atom states $\ket{\Psi_{2}}_{j}$ and 
$\ket{\Psi_{3}}_{j}$.

In fact, the master equation (\ref{masterd}) leads to a set of $(N+1)(N+2)/2$ coupled equations of motion for the populations of the collective dressed 
states. Fortunately, however, an explanation of the enhancement of the CPT effect, seen in Figs.~\ref{fig-4} and \ref{fig-5}, does not require a complete 
solution for the populations of the dressed states. It is enough to consider only the population of the state $\ket{D2}_{N}$ or $\ket{D3}_{N}$. 
Thus, using Eqs.~(\ref{DS}) and (\ref{rab}), we can show that for $N=1$ the stationary population of the state $\ket{D2}_{N}$ is simply equal to 
$\langle R_{22}\rangle$, and for $N>1$ is given by the following expectation value
\begin{eqnarray}
\rho_{D2} &=& \frac{1}{N!}\langle (R_{12}R_{21} + R_{22} - R_{11})R_{11}(R_{11} - 1) \nonumber \\
&\times& (R_{11} - 2)\ldots (R_{11} - N + 2)\rangle, \label{4.5}
\end{eqnarray}
which can be evaluated using the steady-state solution (\ref{SS}).

\begin{figure}[t]
\includegraphics[height=4cm]{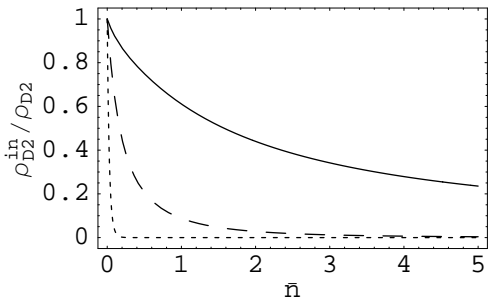}
\caption{\label{fig-7} The ratio $\rho_{D2}^{in}/\rho_{D2}$ as a function of $\bar n_{2} = \bar n_{3} =\bar{n}$ for different numbers 
of atoms: $N=2$ (solid line), $N=4$ (dashed line), $N=20$ (short dashed line).}
\end{figure}

We also calculate the population of the state $\ket{D2}_{N}$ in the case of independent atoms and find
\begin{eqnarray}
\rho_{D2}^{in}=\frac{\Gamma_{2}\left(\Gamma_{1}\right)^{N-1}}{\left(\Gamma_{1} + 2\Gamma_{2}\right)^{N}}. \label{4.6}
\end{eqnarray}
To see the difference between the populations (\ref{4.5}) and (\ref{4.6}), we study the ratio $\rho_{D2}^{in}/\rho_{D2}$.
Figure \ref{fig-7} shows the ratio for different numbers of atoms. For $N=1$, the ratio is equal to one, but for $N>1$ the ratio is smaller than one 
and decreases with $N$. This shows that the population of the collective states is larger than the population of the equivalent states of independent atoms. 
In other words, we may say that the capacity of the collective states is larger than the capacity of the equivalent states of independent atoms. 

The above analysis give us a simple physical interpretation of the collective trapping effect. We may conclude that the improvement of trapping effect by 
multiatom system is due simply to the increased storage capacity of the collective (entangled) states compared with the storage capacity of the equivalent 
states of independent atoms.
\section{Summary}
We have investigated the coherent population trapping effect in a collective system of three level atoms driven by two coherent laser fields and 
simultaneously coupled to the reservoir of a non-zero temperature. The thermal reservoir causes thermal decoherence which affects the trapping effect. 
We have shown that in a single atom there is no perfect CPT when both atomic transitions are affected by thermal decoherence. The perfect CPT may occur 
when only one of the two atomic transitions is affected by thermal decoherence. Extending the analysis to multi-atom systems, we have shown that the 
destructive effect of the thermal decoherence on the CPT can be circumvented by the collective behavior of the atoms. Unlike the case of noninteracting 
atoms in which decoherence processes are independent of the number of atoms, we have found that the collective behavior of the atoms can substantially 
improve the trapping effect destroyed by the thermal decoherence. In the collective atomic system the trapping effect increases with increasing number 
of atoms. If number of atoms is large enough, an almost complete CPT is observed even at high temperatures of the reservoir.
This feature is explained in terms of the semiclassical dressed atom model. We have shown that the improvement of the CPT trapping in the multiatom system 
arises from the presence of collective (entangled) states whose capacity of storage of the atomic population is larger than the corresponding states of 
independent atoms.

\section*{ACKNOWLEDGMENTS}
ZF would like to thank The Max-Planck Institute for hospitality and The University of Queensland for 
the travel support.

{\small $^\star$ On leave from \it{Technical University of Moldova, Physics Department, 
\c{S}tefan Cel Mare Av. 168, MD-2004 Chi\c{s}in\u{a}u, Moldova.}}

\end{document}